\documentclass[reprint,amsmath,amssymb,aps,nofootinbib]{revtex4-2}
\usepackage{graphicx}
\graphicspath{ {images/} }
\usepackage{epstopdf}
\usepackage{hyperref}

\usepackage{color}
\definecolor{green}{rgb}{0,0.5,0}
\usepackage{xcolor}
\usepackage{textcomp}

\begin{document}

\title{Diffusion in multicomponent granular mixtures}
\author{Anna Bodrova}
\date{\today}

\affiliation{Moscow Institute of Electronics and Mathematics, HSE University, 123458, Moscow, Russia}

\begin{abstract}
We investigate diffusion in a polydisperse granular media. We derive mean-squared displacement of granular particles in a polydisperse granular gas in a homogeneous cooling state, containing arbitrary amount of species of different sizes and masses. We investigate both models of constant and time-dependent restitution coefficients and obtain a universal law for the size-dependence of the mean-squared displacement for steep size distributions. 
\end{abstract}

\maketitle

\section{Introduction}

Granular materials are commonly used in nature and technology \cite{GranRev, PhysGranMed, SanPowGr, DryGranMed}. There are numerous examples: stones and sand in the building industry; grains, sugar, salt, and cereals in the food industry; and powders in cosmetic production. Granular dust covers the surface of Mars \cite{mars}, other planets and satellites.  

Diluted granular systems are termed as granular gases \cite{book}. The distance between the particles is larger than their sizes and the total packing fraction of all components usually does not exceed approximately 20\%. Granular gases are components of large interstellar dust clouds, protoplanetary discs and planetary rings in space. They can be obtained in a microgravity environment on planes and rockets \cite{Sperl} by placing granular matter in containers with vibrating \cite{vib1, vib2} or rotating \cite{rot} walls, applying electrostatic \cite{el} or magnetic forces \cite{magn1, magn2}.

The theory of granular gases extends the ideal gas model to include dissipation of particle collisions. Thus, a granular gas represents a fundamental physical system in statistical mechanics. It can be considered a reference model system in granular matter physics \cite{mehta}. Most studies on granular gases have been devoted to one-component granular gas \cite{book}, whereas granular systems are mostly polydisperse. In Saturn's rings the sizes of granular particles range from millimeters to several meters \cite{rings, pnas}.

Because of the dissipative nature of inter-particle collisions, the energy equipartition, valid for equilibrium molecular systems, does not hold for granular mixtures, where each species $k$ has its
own temperature $T_k$ \cite{garzoreview}.
\begin{equation}
\frac32 n_k T_k=\frac{m_k\langle v_k^2\rangle}{2} =\int d {\bf v}_k f_k\left({\bf v}_k, t\right)\frac{m_kv_k^2}{2}
\end{equation}
Here, $m_k$ is the mass of the granular species,  ${\bf v}_k$ is its velocity, $f\left({\bf v}_k,t\right)$ is the
velocity distribution function, which quantifies the number of particles in the system of type $k$ with velocity $\textbf{v}_k$ at time $t$ and $n_k$ is the number density of the $k$-th species of the  granular fluid:
\begin{equation}
n_k= \int f_k\left({\bf v}_k,t\right) d {\bf v}_k .
\end{equation}
The total number density is equal to
\begin{equation}
n=\sum_i n_i
\end{equation}
The velocity distribution function  of species $k$ is assumed to be Maxwellian 
\begin{equation}\label{fmax}
f_k\left(\textbf{v}_k,t\right)=n_k\left(\frac{m_k}{2\pi T_k}\right)^{3/2}\exp\left(-\frac{m_kv_k^2}{2T_k}\right)
\end{equation}
In force-free granular gases the granular temperatures decay owing to dissipative collisions, leading to anomalous subdiffusive motion of the granular particles \cite{book}.

Anomalous diffusion represents the motion of particles, characterised by the non-linear dependence of the mean-squared displacement (MSD) on time \cite{ralf, sok, eli, georges, franosh}:
\begin{equation}
\left\langle R^2(t) \right\rangle \sim t^{\alpha}
\end{equation}
where $\alpha\ne 1$, while $0<\alpha<1$ corresponds to subdiffusion \cite{jasmin} and $\alpha>1$ corresponds to superdiffusion. The case $\alpha = 2$
describes the ballistic motion, and the cases with $\alpha > 2$ are
termed superballistic or hyperdiffusive.
The logarithmic time-dependence of MSD corresponds to ultraslow motion \cite{ultraslow, jeon}:
\begin{equation}
\left\langle R^2(t) \right\rangle \sim \log t
\end{equation}
Depending on the restitution coefficient model, the motion of particles in a force-free cooling unicomponent granular gas may be either ultraslow or subdiffusive \cite{annapccp}. The effect of the roughness of the granular particles on the diffusion coefficient was also considered \cite{annaselfdif}. The MSD of granular intruders has been investigated in binary granular mixtures \cite{garzointruder, annaprl} and granular suspensions \cite{garzointruder23}. However, diffusion in multicomponent granular mixtures has not been studied previously. Considering that granular systems are usually highly polydisperse, one needs to fill this important gap, which is the subject of investigation in the current article. In the next Section II we describe the model of both constant and velocity-dependent restitution coefficients. In Sec. III we review the derivation of the granular temperature \cite{haff}, velocity correlation time \cite{book} and MSD in a unicomponent granular gas with a constant restitution coefficient. In Sec. IV we discuss the known results for the cooling rates in polydisperse granular mixtures. In Sec. V we derive the partial MSDs in a polydisperse granular mixture, which is the main contribution of our paper. Finally, in the last Sec. VI  we summarize our findings. 

\section{Models of the restitution coefficient}

The dissipative nature of inter-particle collisions is quantified by the so-called restitution coefficient $\varepsilon$, see e.g.   \cite{GranRev,book}:
\begin{equation}
\label{rc} \varepsilon = \left|\frac{\left({\bf v}^{\,\prime}_{ki} \cdot {\bf e}\right)}{\left({\bf v}_{ki} \cdot {\bf
e}\right)}\right| \, .
\end{equation}
Here  ${\bf v}_{ki}={\bf v}_{k}-{\bf
v}_{i}$ and ${\bf v}^{\,\prime}_{ki}={\bf v}_{k}^{\,\prime}-{\bf v}_{i}^{\,\prime}$ are the relative velocities of particles of masses $m_k$ and $m_i$ before and after a  collision, respectively, and ${\bf e}$ is a  unit vector directed along the inter-center vector  at the collision instant. 
The post-collision velocities  ${\bf v}_{k}^{\,\prime}$ and ${\bf v}_{i}^{\,\prime}$ are  related to the pre-collision velocities ${\bf v}_{k}$ and ${\bf v}_{i}$ as follows, \cite{book}:
\begin{equation}\label{v1v2} {\bf v}_{k/i}^{\,\prime} = {\bf v}_{k/i} \mp  \frac{m_{\rm eff}}{m_{k/i}}\left(1+\varepsilon\right)({\bf v}_{ki} \cdot {\bf e}){\bf e} \, .\end{equation}
Here $m_{\rm eff}=m_im_k/\left(m_i+m_k\right)$ is the effective mass of colliding particles.
The restitution coefficient $0 \leq \varepsilon <1$ indicates that the
after-collisional relative velocity is smaller than the pre-collisional velocity because the mechanical energy is transformed 
into the internal degrees of freedom of the particles. In oblique collisions, the restitution coefficient may also attain negative values \cite{Saitoh}.

Several models have been proposed for the restitution coefficient. The simplest assumption is that the restitution coefficient is equal to a constant value \cite{book}. It is easy to implement in the analytical calculations and can be considered as a basic reference model. However, this model contradicts experiments, showing a clear dependence of the restitution coefficient on impact velocity \cite{rings}. With a decrease in the relative velocity, the restitution coefficient tends to unity, and the collisions become more elastic.

The simplest, but still rigorous, first-principle model of inelastic collisions considers the
viscoelastic properties of the particle material. This results in the corresponding inter-particle forces
\cite{bshp96,goldobin} and eventually, in the restitution coefficient for viscoelastic particles \cite{rpbs99,delayed}
\begin{eqnarray}\label{epsx} 
&&\varepsilon_{ki} = 1 + \sum^{20}_{j=1} h_j \left( A \kappa_{ki}^{2/5} \right)^{j/2} \left |\left
({\bf{v}}_{ki}\cdot\textbf{e}\right )\right |^{j/10}.
\end{eqnarray}
Here, $h_k$ are numerical coefficients, $\kappa$ and $A$ characterize the elastic and dissipative properties of the particle material, respectively. The viscoelastic model agrees well with the experimental data when the impact velocity is not very high \cite{rpbs99}. The dissipative constant $A$ quantifies the viscous properties of the particle material \cite{bshp96,goldobin}:
\begin{equation}
\label{A} A =
\frac{1}{Y}\frac{\left(1+\nu\right)}{\left(1-\nu\right)}\left(\frac43\eta_1\left(1-\nu+\nu^2\right)+\eta_2\left(1-2\nu\right)^2\right)
\end{equation}
where ${\eta}_1$ and ${\eta}_2$ are viscosity coefficients. 
$\kappa_{ij}$ is a function of Young's modulus $Y$, Poisson's ratio $\nu$, mass, and size of particles \cite{book,bshp96,rpbs99}.
\begin{eqnarray}
\kappa_{ki}=\frac{1}{\sqrt{2}} \left (\frac{3}{2}\right )^{3/2}\frac{Y}{1-{\nu}^2}\frac{\sqrt{\sigma_{\rm eff}}}{m_{\rm eff}}
\end{eqnarray}
The effective diameter of colliding particles with diameters $\sigma_i$ and $\sigma_j$ is
\begin{equation}
\sigma_{\rm eff}=\frac{\sigma_i\sigma_j}{\sigma_i+\sigma_j}
\end{equation}

\section{Unicomponent granular gas with constant restitution coefficient}

Let us first consider a monodisperse granular gas. Taking into account that the displacement of the particle may be expressed as the integral of its velocity, $\textbf{R}=\int_0^t\textbf {v}(t^{\prime})dt^{\prime}$, the MSD can be written in terms of the velocity correlation function:
\begin{equation}
\left< R^2(t)\right>=\int_0^tdt_1 \int_0^t dt_2 \left<\textbf{v}_1(t_1)\textbf{v}_1(t_2)\right>
\end{equation}

The evolution of the velocity can be described in terms of the pseudo-Liouville operator \cite{book}:
\begin{eqnarray}\label{Liu}
\frac{d\textbf{v}_1}{dt}=\mathcal{L} \textbf{v}_1
\end{eqnarray}
The pseudo-Liouville operator may be expressed trough the sum of the free streaming component $\mathcal{L}_0=\mathbf{v}_1\nabla_{\mathbf{r}_1}$ and binary collision operators \cite{book}:
\begin{eqnarray}
\mathcal{L} = \mathcal{L}_0+\sum_{i>1}\hat{T}_{1i}\end{eqnarray}
The binary collision operator describes the interactions between particles $i$ and $j$
\begin{eqnarray}\nonumber
\hat{T}_{1i} = \sigma^2\int d\textbf{e} \Theta\left(-\textbf{v}_{1i}\cdot\textbf{e}\right)|\textbf{v}_{1i}\cdot\textbf{e}|\delta\left(\textbf{r}_{1i}-\sigma\textbf{e}\right)\left(\hat{b}_{1i}-1\right)\\\label{eq:Tqi}
\end{eqnarray}
Operator $\hat{b}_{1i}$ replaces the pre-collision velocities with the post-collision velocities, using Eqs.~(\ref{v1v2}). The Heaviside step-function $\Theta(x)$ selects the approaching particles and the $\delta$-function determines the unit vector $\textbf{e}$, which specifies the collision. 

Due to the assumption of the molecular chaos the velocities of particles are not correlated. The velocities of particles change at the collision instants without memory, which corresponds to the Markov process. The time-correlation function of a Markov process is an exponentially decaying function \cite{resibois}. Therefore the velocity correlation function, corresponding to the adiabatic approximation, takes the form:
\begin{equation}\label{velcorad}
\left\langle \textbf{v}_1(t^{\prime})\cdot \textbf{v}_1(t) \right\rangle = \frac{3 T_1(t^{\prime})}{m_1}\exp\left(-\frac{|t-t^{\prime}|}{\tau_{v,\,\rm ad}(t^{\prime})}\right)
\end{equation}
In the adiabatic approximation it is assumed that the prefactor changes much slower than the exponential term. The adiabatic velocity correlation time equals to the initial slope of the time-correlation function or may be presented as the derivative at $t=t^{\prime}+\epsilon$ in the limit $\epsilon\to 0^+$:
\begin{equation}
\tau_{v,\,\rm ad}^{-1}(t)=-\frac{m_1}{3T_1}\lim_{t\to t^{\prime}+0^+}\frac{d}{dt}\left\langle \textbf{v}_1(t^{\prime})\cdot \textbf{v}_1(t) \right\rangle
\end{equation}
In order to derive $\tau_{v,\,\rm ad}^{-1}(t)$ we perform formal integration of Eq.~(\ref{Liu}) and obtain the velocity time-correlation function in the form \cite{book}:
\begin{equation}
\left\langle \textbf{v}_1(t^{\prime})\cdot \textbf{v}_1(t) \right\rangle = \left\langle \textbf{v}_1(t^{\prime}) \exp\left(\mathcal{L}(t-t^{\prime})\right) \textbf{v}_1(t^{\prime}) \right\rangle
\end{equation}
Taking the derivative with respect to time $t$, one can express the adiabatic velocity correlation time in terms of the pseudo-Liouville operator:
\begin{equation}\tau_{v,\,\rm ad}^{-1}(t)=-\frac{m_1}{3T_1} \left\langle\textbf{v}_1 \mathcal{L} \textbf{v}_1 \right\rangle\end{equation} 
In order to perform the averaging over the two-particle correlation function $f_2\left(\textbf{r}_1,\textbf{r}_2,\textbf{v}_1, \textbf{v}_2\right)$, we use the hypothesis of molecular chaos, which allows us to represent $f_2\left(\textbf{r}_1,\textbf{r}_2,\textbf{v}_1, \textbf{v}_2\right)$ as a product of one-component distribution functions:
\begin{equation}
f_2\left(\textbf{r}_1,\textbf{r}_2,\textbf{v}_1, \textbf{v}_2\right)=g_2(\textbf{r}_{12})f_1\left({\bf v}_1, t\right)f_1\left({\bf v}_2, t\right)
\end{equation}
where the coordinate part factorizes from the velocity part, $\textbf{r}_{12}=\textbf{r}_1-\textbf{r}_2$ and $g_2\left(\textbf{r}_{12}\right)$ is the contact value of the pair correlation function \cite{book}. This may be assumed to be equal to unity for dilute systems. Using $f_1\left({\bf v}_1, t\right)$ in the form of Maxwellian distribution (Eq.~\ref{fmax}) and performing the averaging one obtains the inverse adiabatic velocity correlation time:
\begin{equation}\label{tauv1}
\tau_{v,\,\rm ad}^{-1}(t)=\frac{8}{3}n\sigma^2g_2(\sigma)\sqrt{\frac{\pi T}{m}}\frac{1+\varepsilon}{2}
\end{equation}
Beyond the adiabatic approximation the velocity correlation function of a particle takes the following form
\begin{equation}\label{velcor}
\left\langle \textbf{v}_1(t^{\prime})\cdot \textbf{v}_1(t) \right\rangle = 3\sqrt{\frac{T_1(t^{\prime})T_1(t)}{m_1}}\exp\left(-\frac{|t-t^{\prime}|}{\tau_v(t^{\prime})}\right)
\end{equation}
and the inverse velocity correlation time is equal to
\begin{equation}\label{tauxi}
\tau_{v}^{-1}(t)=\tau_{v,\,\rm ad}^{-1}(t)-\frac{1}{2}\xi(t)
\end{equation}
Here $\xi(t)$ is the cooling rate, quantifying the decay of the granular temperatures in the granular gas in homogeneous cooling state \cite{book}:
\begin{equation}\label{difftemp}
\xi(t) = -\frac{1}{T}\frac{dT(t)}{dt}
\end{equation}
In a one-component granular gas it reads
\begin{equation}\label{xit}
\xi(t) = \frac{1-\varepsilon^2}{3}\tau_c^{-1}(t)
\end{equation}
Differential equation (Eq.~\ref{difftemp}) with cooling rate (Eq.~\ref{xit}) may be solved explicitly, and the temperature obeys Haff's law \cite{haff}:
\begin{equation}
\frac{T(t)}{T(0)} = \left(1+\frac{t}{\tau_0}\right)^{-2}
\end{equation}
The inverse characteristic time of the granular temperature decay was equal to half of the cooling rate at the initial time
\begin{equation}\label{tau0}
\tau_0^{-1} = \frac{1-\varepsilon^2}{6}\tau_c^{-1}(0)=\frac{\xi(0)}{2}
\end{equation}
The inverse mean collision time is
\begin{equation}\label{tauc}
\tau_c^{-1}(t)=4n\sigma^2g_2(\sigma)\sqrt{\frac{\pi T}{m}}
\end{equation}
Introducing Eqs.~(\ref{xit}) and (\ref{tauv1}) into Eq.~(\ref{tauxi}) yields
\begin{equation}
\tau_{v}^{-1}(t)=\tau_E^{-1}(t)\frac{\left(1+\varepsilon\right)^2}{4} = \tau_c^{-1}(t)\frac{\left(1+\varepsilon\right)^2}{6}
\end{equation}
For elastic particles with $\varepsilon = 1$ it becomes equal to the Enskog relaxation time
\begin{equation}\label{taue}
\tau_E^{-1}(t)=\frac{8}{3}n\sigma^2g_2(\sigma)\sqrt{\frac{\pi T}{m}}=\frac{2}{3}\tau_c^{-1}(t)
\end{equation}
The time-dependent diffusion coefficient may be derived as
\begin{equation}\label{diff}
D(t) = \frac{T(t)\tau_v(t)}{m}
\end{equation}
The MSD of a one-component granular gas with a constant restitution coefficient has the following form \cite{annapccp}:
\begin{eqnarray}\nonumber
&&\left< R^2(t)\right>=6D(0)\tau_0\log\left(1+\frac{t}{\tau_0}\right)\\
&&+6D(0)\tau_v(0)\left(\left(1+\frac{t}{\tau_0}\right)^{-\beta}\!-1\right)
\end{eqnarray}
Here 
\begin{equation}
\beta = \frac{\tau_0}{\tau_v(0)} = \frac{\left(1+\varepsilon\right)^2}{1-\varepsilon^2}
\end{equation}
At short times $t\ll\tau_0$ the MSD has ballistic behavior $\left<R^2(t)\right>\sim t^{2}$, at long times it has logarithmic time-dependence, $\left<R^2(t)\right>\sim \log t$. The MSD of particles with a viscoelastic restitution coefficient scales according to $\left< R^2(t)\right>\sim t^{1/6}$ at long times.

\section{Granular temperatures in a polydisperse mixture}

The distribution of granular temperatures in granular mixtures has been studied for both constant \cite{lev} and viscoelastic \cite{Os} restitution coefficients as well as for granular particles immersed in molecular gas \cite{Osgas}. Let us now consider a mixture of $N$ species. The evolution of partial granular temperatures in a mixture occurs according to the following system of differential equations \cite{lev, zippelius, hrenya}:
\begin{eqnarray}
\frac{dT_k}{dt}=-T_k\xi_k\,\\ k = 1,...,N
\end{eqnarray}
The cooling rate is equal to the sum
\begin{equation}
\xi_k=\sum_{i=1}^N\xi_{ki}
\end{equation}
For the constant restitution coefficient the cooling rate $\xi_{ki}$, quantifying the decrease in the granular temperature of species of mass $m_k$ due to collisions with species of mass $m_i$ is given by the following expression \cite{lev, zippelius, hrenya}:
\begin{eqnarray}\nonumber
&&\xi_{ki}(t) =\frac{8}{3}\sqrt{2\pi}n_i\sigma_{ki}^{2}g_{2}(\sigma_{ki})
\left(\frac{T_km_i+T_im_k}{m_im_k}\right)^{1/2}\!\left(1+ \varepsilon_{ki}\right)\\
&& \times\left(\frac{m_i}{m_i+m_k}\right)\left[1-\frac{1}{2}\left(1+ \varepsilon_{ki}
\right)\frac{T_im_k+T_km_i}{T_k\left(m_i+m_k\right)} \right] \label{xikconst}.
\end{eqnarray}
Here $\sigma_{ki}=\left(\sigma_k+\sigma_i\right)/2$. 
The cooling rates for the constant restitution coefficient became equal after a short relaxation time, leading to constant ratio of granular temperatures during the evolution of the system.

In the case of binary mixture with $n_k\ll n_i$, $m_k\gg m_i$, $\varepsilon_{ii}\to 1$ the temperature ratio may be found explicitly \cite{book, garzoreview, brey}:
\begin{eqnarray}
\frac{T_k}{T_i}&=&\frac{1+\varepsilon_{ik}}{2\left(1-b\right)}\\
b &=& \frac{1}{2\sqrt{2}}\frac{1-\varepsilon_{ii}^2}{1+\varepsilon_{ik}}\frac{m_k}{m_i}\frac{\sigma_{ik}^2}{\sigma_k^2}\frac{g_2(\sigma_{ik})}{g_2(\sigma_k)}
\end{eqnarray}

Granular temperatures in a viscoelastic one-component granular gas scale at long times according to \cite{book}
\begin{equation}
T_i(t)\sim t^{-5/3}
\end{equation}
and the cooling rates attain the form \cite{Os}
\begin{eqnarray}\nonumber
&&\!\xi_{ki}(t) =
\frac{16}{3}\sqrt{2\pi}n_i\sigma_{ki}^{2}g_{2}(\sigma_{ik})
\left(\frac{T_km_i+T_im_k}{m_im_k}\right)^{1/2}\!\left(\frac{m_i}{m_i+m_k}\right)\\\nonumber
&&\!\times\left[1-\frac{T_km_i+T_im_k}{T_k\left(m_i+m_k\right)}+\sum_{n}B_n\left(h_n-\frac12\frac{T_km_i+T_im_k}{T_k\left(m_i+m_k\right)}A_n\right) \right]\\
\label{xikvisc}
\end{eqnarray}
where $A_n=4h_n+\sum_{j+k=n}h_jh_k$ are pure numbers and
\begin{equation}
B_n(t)\!=\left(A\kappa_{ki}^{2/5}\right)^{\frac{n}{2}}\!\left(2\frac{T_km_i+T_im_k}{m_im_k}\right)^{n/20}\!\left(\frac{\left(20+n\right)n}{800}\right)\!\Gamma\left(\frac{n}{20}\right)
\end{equation}
where $\Gamma\left(x\right)$ denotes the gamma-function. The ratio of granular temperatures does not remain constant as compared to the case of constant restitution coefficient, and the system tends to equipartition with the passage of time \cite{Os}.

\section{Mean-squared displacement in a polydisperse granular mixture}


In a mixture of $N$ species the total MSD can be expressed through partial MSDs
\begin{equation}\label{MSD}
\left< R^2(t)\right>=\frac{1}{n}\sum_{k=1}^N n_k\left< R^2_k(t)\right>
\end{equation}
Let us derive the partial MSD of species $k$. The evolution of velocity $\textbf{v}_k$ may also be described in terms of the pseudo-Liouville operator as for the one-component granular gas
\begin{eqnarray}
\frac{d\textbf{v}_k}{dt}=\mathcal{L} \textbf{v}_k
\end{eqnarray}
The Lioville operator takes the form
\begin{eqnarray}
\mathcal{L} = \mathcal{L}_0+\sum_{i=1}^NN_i\hat{T}_{ki}\end{eqnarray}
Here $N_i$ is the number of particles of species $i$, $\hat{T}_{ki}$ is the collisional operator describing the interactions between particles of species $i$ and $k$, respectively:
\begin{eqnarray}
\hat{T}_{ki} = \sigma^2\int d\textbf{e} \Theta\left(-\textbf{v}_{ki}\cdot\textbf{e}\right)|\textbf{v}_{ki}\cdot\textbf{e}|\delta\left(\textbf{r}_{ki}-\sigma\textbf{e}\right)\left(\hat{b}_{ki}-1\right)
\end{eqnarray}
Taking into account that $\textbf{R}_k=\int_0^t\textbf {v}_k(t^{\prime})dt^{\prime}$, the MSD can be written in terms of the velocity correlation function:
\begin{equation}
\left< R^2_k(t)\right>=\int_0^tdt_1 \int_0^t dt_2 \left<\textbf{v}_k(t_1)\textbf{v}_k(t_2)\right>
\end{equation}
Introducing the reduced time $\tau_k$  according to
\begin{equation}
d\tau_k=dt\sqrt{T_k(t)/T_k(0)}\tau_c^{-1}(0)\,,
\end{equation}
we can write down the exponential correlation function of the reduced velocities $\textbf{c}_k=\textbf{v}_k\sqrt{m_k/(2T_k)}$ \cite{book}:
\begin{equation}
\left<\textbf{c}_k\left(\tau_k(t_1)\right)\textbf{c}_k\left(\tau_k(t_2)\right)\right>=\frac{3}{2}\exp\left[-\frac{\tau_k(t_2)-\tau_k(t_1)}{\hat{\tau}_{v,k}(t_1)}\right]
\end{equation}
and express the partial MSD $\left< R^2_k(t)\right>$ as
\begin{equation}\label{R2}
\left< R^2_k(t)\right>=6\int_0^tdt_1 D_k(t_1)\left[1-\exp \left(-\frac{\tau_k(t)-\tau_k(t_1)}{\hat{\tau}_{v,k}(t_1)}\right)\right]
\end{equation}

\begin{figure}\centerline{\includegraphics[width=0.95\columnwidth]{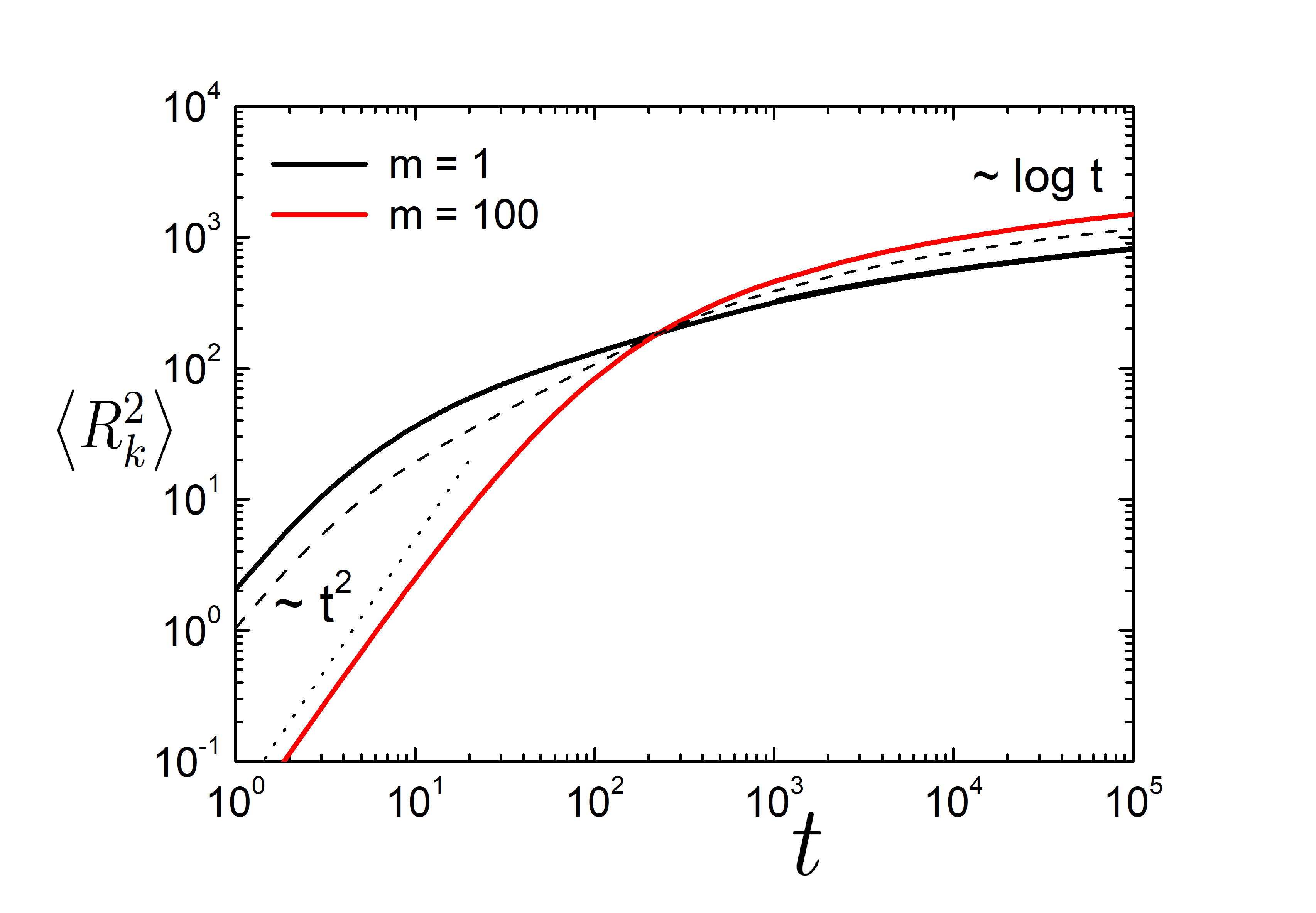}}\caption{Time-dependence of partial MSDs in a binary granular mixture for particles, colliding with a constant restitution coefficient $\varepsilon=0.5$. The partial number densities of particles are equal: $n_1=n_2=0.1$. The masses of species are $m_1=1$, $m_{2}=100$, the diameters $\sigma_1=1$, $\sigma_{2}=1$. At short time the particles move along ballistic trajectories $\left< R^2_k(t)\right>\sim t^2$, at long times the particles perform ultraslow diffusion $\left< R^2_k(t)\right>\sim \log t$ (Dotted lines show the slope). Dashed line corresponds to the total MSD $\left< R^2(t)\right>$ (Eq.~(\ref{MSD})). } \label{Gm1m100}
\end{figure}

\begin{figure}\centerline{\includegraphics[width=0.95\columnwidth]{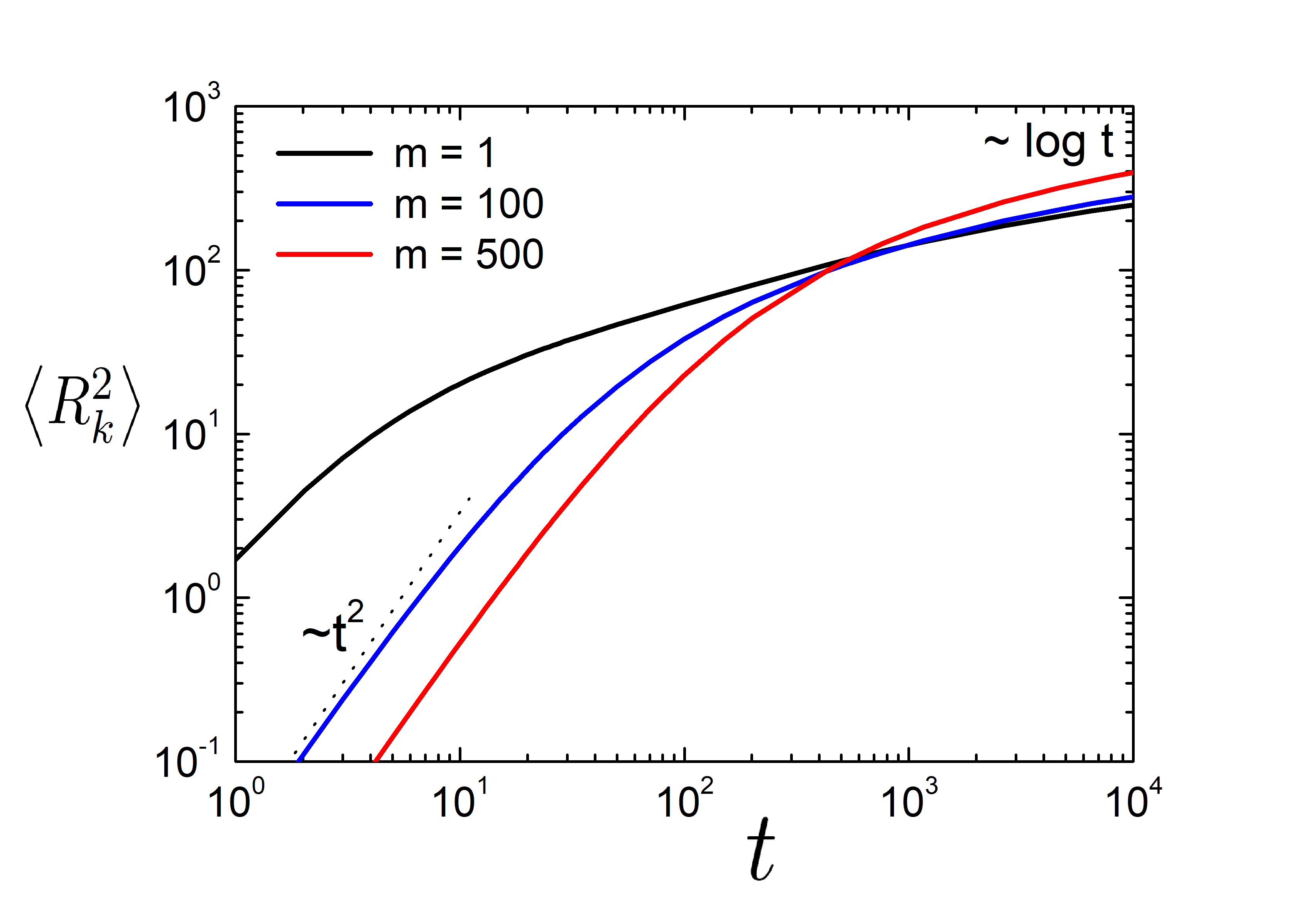}}\caption{Time-dependence of partial MSDs in a tertiary granular mixture for particles, colliding with a constant restitution coefficient $\varepsilon=0.5$ (solid lines). The number densities of particles are equal: $n_1=n_2=n_3=0.1$. The masses of species are $m_1=1$, $m_{2}=100$, $m_{3}=500$, the diameters $\sigma_1=1$, $\sigma_{2}=1$, $\sigma_{3}=1$. } \label{Gm1m100m500}\end{figure}

Here the reduced velocity correlation time is
\begin{equation}\label{hattau}
\hat{\tau}_{v,k}(t)=\tau_{v,k}(t)\sqrt{\frac{T_k(t)}{T_k(0)}}\tau_c^{-1}(0)
\end{equation}
The partial diffusion coefficient of species $k$ may be calculated according to
\begin{equation}\label{diffk}
D_k(t)=\frac{T_k(t)\tau_{v,k}(t)}{m_k}
\end{equation}
The inverse velocity correlation time is given by the sum
\begin{equation}\label{tausum}
\tau_{v,k}^{-1}(t)=\sum_{i=1}^N \tau_{v,ki}^{-1}(t)
\end{equation}
where
\begin{equation}\label{tauadxi}
\tau_{v,ki}^{-1}(t)=\tau_{v,ki,\,\rm ad}^{-1}(t)-\frac12\xi_{ki}(t)
\end{equation}
The adiabatic velocity correlation time is calculated as follows:
\begin{eqnarray}
\tau_{v,k\,\rm ad}^{-1}(t)=\sum_{i=1}^N \tau_{v,ki,\,\rm ad}^{-1}(t)\\
\tau_{v,ki\,\rm ad}^{-1}(t)=-\frac{N_im_k}{3T_k} \left\langle\textbf{v}_k \hat{T}_{ki} \textbf{v}_k \right\rangle
\end{eqnarray}

\subsection{Constant restitution coefficient}

For $\varepsilon = \rm const$ we get after performing the averaging 
\begin{eqnarray}\nonumber
&&\tau_{v,ki,\,\rm ad}^{-1}(t)=\frac{8\sqrt{2\pi}}{3}n_i\sigma_{ki}^{2}g_{2}(\sigma_{ki})\frac{m_i}{m_i+m_k}\\&&\times\left(\frac{T_km_i+T_im_k}{m_im_k}\right)^{1/2}\frac{1+\varepsilon}{2}\label{tauvconstad}
\end{eqnarray}
Introducing Eqs.~(\ref{xikconst}) and (\ref{tauvconstad}) into Eq.~(\ref{tauadxi}) yields
\begin{eqnarray}
\nonumber
&&\tau_{v,ki}^{-1}(t)=\frac{8\sqrt{2\pi}}{3}n_i\sigma_{ki}^{2}g_{2}(\sigma_{ki})\frac{m_i}{m_i+m_k}\frac{T_km_i+T_im_k}{T_k\left(m_i+m_k\right)}\\&&\times\left(\frac{T_km_i+T_im_k}{m_im_k}\right)^{1/2}\frac{\left(1+\varepsilon\right)^2}{4}\label{tauvconst}
\end{eqnarray}

\begin{figure}\centerline{\includegraphics[width=0.95\columnwidth]{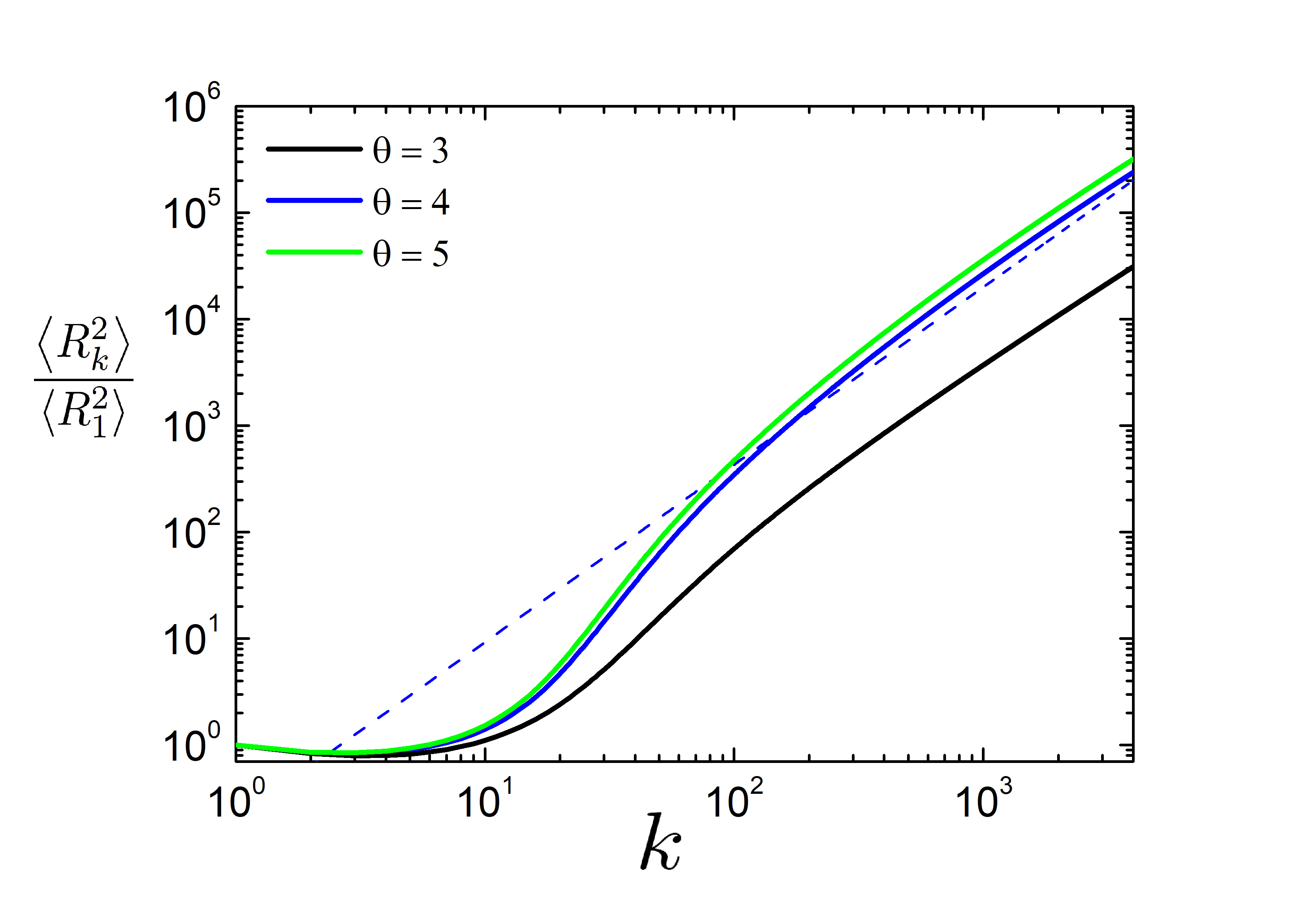}}\caption{Fraction of partial MSDs $\left<R_k^2\right>/\left<R_1^2\right>$ for different values of $k = m_k/m_1$ in a mixture of granular particles with number density $n_k = n_1k^{-\theta}$, $\theta=3,4,5$, $n_1 = 0.1$ number of species $N = 4000$, time $t = 100000$ and constant restitution coefficient $\varepsilon = 0.5$. Solid line corresponds to the result of numerical integration of Eq.~\ref{R2}, thin dashed line shows the scaling $\left< R^2_k\right>\sim k^{5/3}$.} \label{GRk}\end{figure}

Introducing Eq.~(\ref{hattau}-\ref{diffk}), (\ref{tauvconst}) into Eq.~(\ref{R2}) and performing numerical integration, we can obtain the partial MSDs in a granular mixture for particles, colliding with a constant restitution coefficient $\varepsilon = \rm const$. The results for the binary and tertiary granular mixtures with equal amounts of species of each mass are shown in Figs. \ref{Gm1m100} and \ref{Gm1m100m500}, respectively. 
In binary mixtures, at short times, particles with lighter masses also move faster. However, at longer times the motion of particles with mass $m=1$ becomes slower than that of particles with mass $m=100$, 
(Fig.~\ref{Gm1m100}). The trajectories of massive particles are slightly affected by collisions with much lighter particles. In tertiary mixtures, the fastest particles have the largest mass $m=500$ (Fig.~\ref{Gm1m100m500}).


Now we investigate mixtures with a large amount of species $N\gg 1$. If the size distribution is steep enough ($\theta>2$), the granular temperature distribution scales according to $T_k\sim k^{5/3}$ \cite{lev}. Then $\tau_{v,ki}^{-1}\sim i^{2-\theta}/k$ at $k\gg i$. Replacing in Eq.~(\ref{tausum}) the sum by the integral, we obtain that for $N\to\infty$ the integral becomes finite if $\theta>3$, and $\tau_{v,k}(t)\sim k$. Therefore, $D_k(t)\sim k^{5/3}$, and for mean-squared displacements we obtain the same scaling law as for granular temperatures: $\left< R^2_k\right>\sim k^{5/3}$ for $k\gg 1$, $t\gg\tau_0$. The ratio $\left< R^2_k\right>/\left< R^2_1\right>$ is plotted at Fig.~\ref{GRk} for different values of parameter $\theta$. At $k\gg 1$ a nice agreement with scaling is observed.

\begin{figure}\centerline{\includegraphics[width=0.95\columnwidth]{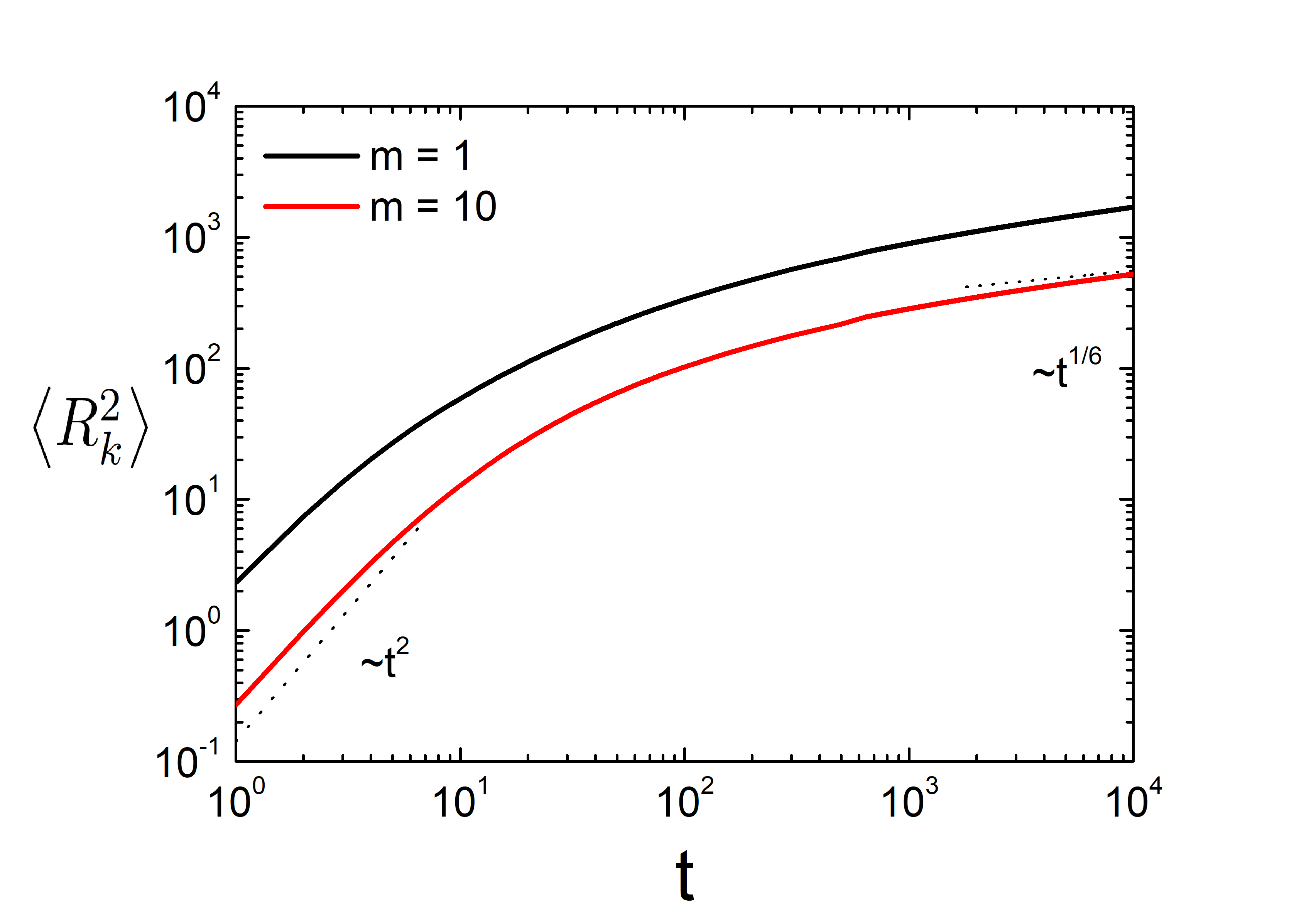}}\caption{MSD in a binary granular mixture with time-dependent restitution coefficient, $A\kappa^{2/5}=0.1$, $m_1=1$, $m_2=10$, $n_1=0.1$, $n_2=0.02$.} \label{GR2visco}\end{figure}

\subsection{Velocity-dependent restitution coefficient}

The inverse adiabatic velocity correlation time for particles with viscoelastic restitution coefficient given by Eq.~(\ref{epsx}) can be analogously derived.
\begin{eqnarray}\nonumber
&&\tau_{v,ki,\,\rm ad}^{-1}(t)=\frac{8\sqrt{2\pi}}{3}n_i\sigma_{ki}^{2}g_{2}(\sigma_{ik})\frac{m_i}{\left(m_i+m_k\right)^2}\\&&\times\left(\frac{T_km_i+T_im_k}{m_im_k}\right)^{1/2}\left(1+\sum_{i}h_iB_i\right)\label{tauvvad}
\end{eqnarray}

The full velocity correlation time reads
\begin{eqnarray}\nonumber
&&\tau_{v,ki}^{-1}(t)=\frac{8\sqrt{2\pi}}{3}n_i\sigma_{ki}^{2}g_{2}(\sigma_{ik})m_i\frac{T_km_i+T_im_k}{T_k\left(m_i+m_k\right)^2}\\&&\times
\left(\frac{T_km_i+T_im_k}{m_im_k}\right)^{1/2}\left(1+\frac{1}{2}\sum_{i}A_iB_i\right)\label{tauvv}
\end{eqnarray}

Introducing Eq.~(\ref{hattau}-\ref{diffk}), (\ref{tauvv}) into Eq.~(\ref{R2}), we can derive partial MSDs $\left< R^2_k(t)\right>$ for particles colliding with the velocity-dependent restitution coefficient. We show the corresponding values for a binary granular mixture at Fig.~\ref{GR2visco}. At short time the particles move along ballistic trajectories,  $\left< R^2_k\right>\sim t^2$, at long times $\left< R^2_k\right>\sim t^{1/6}$.


\section{Conclusions}

We investigated the diffusion in granular mixtures and derived expression for mean-squared displacement for granular mixtures with arbitrary number of different species. We obtained generalized scaling law for partial MSDs in polydisperse granular mixtures with discrete masses $m_k$ and steep number densities $n_k\sim k^{-\theta}$ with $\theta>3$. Our results may be helpful for industrial applications in different branches and investigation of motion of particles in planetary rings and other astrophysical objects.

\section{Acknowledgements}

A.B. thanks A. Osinsky, N. V. Brilliantov, N.G. Iroshnikov and R. Metzler for fruitful discussions.

\end{document}